\documentclass[twocolumn,superscriptaddress,prl]{revtex4}
\usepackage{amsmath}

\usepackage{graphicx}
\usepackage{dcolumn}
\usepackage{bm}
\usepackage{xcolor} 
\usepackage[normalem]{ulem} 
\newcommand{\ML}{}
\newcommand{\MLR}{}
\newcommand{\stkout}[1]{}
\newcommand{\stkoutR}[1]{}

\newcommand{\kT}{k_{\rm B}T}
\newcommand{\md}{\mathrm{d}}
\newcommand{\st}{\boldsymbol{\phi}} 
\newcommand{\St}{\boldsymbol{\Phi}} 
\newcommand{\env}{\boldsymbol{\psi}} 
\newcommand{\Env}{\boldsymbol{\Psi}} 
\newcommand{\org}{s_{\rm{-}}} 
\newcommand{\Org}{S_{\rm{-}}} 
\newcommand{\out}{s_{\rm{+}}} 
\newcommand{\Out}{S_{\rm{+}}} 
\newcommand{\subens}{\boldsymbol{s}} 
\newcommand{\Subens}{\boldsymbol{S}} 
\newcommand{\X}{X}
\newcommand{\Y}{\boldsymbol{Y}}
\newcommand{\x}{x}
\newcommand{\y}{\boldsymbol{y}}
\newcommand{\TPE}{\rm{R}} 

\begin{document}

\preprint{APS/123-QED}

\title{Information Thermodynamics of the Transition-Path Ensemble}

\author{Miranda D.\ Louwerse}
\email{mdlouwer@sfu.ca}
\affiliation{%
 Department of Chemistry, Simon Fraser University, Burnaby, British Columbia V5A1S6, Canada 
}%
\author{David A.\ Sivak}%
 \email{dsivak@sfu.ca}
\affiliation{%
 Department of Physics, Simon Fraser University, Burnaby, British Columbia V5A1S6, Canada 
}%

\date{\today}

\begin{abstract}
The reaction coordinate describing a transition between reactant and product is a fundamental concept in the theory of chemical reactions. Within transition-path theory, a quantitative definition of the reaction coordinate is found in the committor, which is the probability that a trajectory initiated from a given microstate first reaches the product before the reactant. Here we develop an information-theoretic origin for the committor and show how selecting transition paths from \stkoutR{the equilibrium ensemble} {\MLR a long ergodic equilibrium trajectory} induces entropy production which exactly equals the information that system dynamics provide about the reactivity of trajectories. This equality of entropy production and dynamical information generation also holds at the level of arbitrary individual coordinates, providing parallel measures of the coordinate's relevance to the reaction, each of which is maximized by the committor. 
\end{abstract}

\maketitle

Understanding the mechanism for a transition between metastable states of a system is of fundamental interest to the natural sciences. Reaction theories seek to derive the rate constant from underlying system dynamics and have led to increased insight into the reaction mechanism, the sequence of elementary steps by which a reaction occurs. A notable example is transition-state theory and its extensions~\cite{Eyring1935,Evans1935,Kramers1940}, which conceptualize the activated complex (or transition-state species) as a key dynamical intermediate and makes use of its properties (e.g., free energy relative to the reactant) to derive an approximate rate constant for large classes of reactions. The transition state is one identifiable state along the reaction coordinate, a one-dimensional collective variable that preserves all quantitative and qualitative aspects of a reaction under projection of the multidimensional dynamics~\cite{Peters2016,Bolhuis2015}.

Motivated by rare-event sampling methods~\cite{Dellago1998}, transition-path theory~\cite{E2006} was developed to quantitatively describe the entire reaction and determine its rate constant, without assumptions of metastability for the reactant and product or any specific details of the reaction mechanism (e.g., the presence of a single transition state). This statistical description relies on the definition of the committor function $q_{\st}$ (also called the commitment or splitting probability), the probability that a trajectory initiated from microstate $\st$ reaches the product before returning to the reactant. The committor maps the state space onto the interval $q_{\st} \in [0,1]$ and has been called the ``true'' or ``ideal'' one-dimensional reaction coordinate~\cite{Peters2016,E2010,Li2014,Peters2013,Banushkina2016}. The committor allows calculation of the reaction rate from a one-dimensional description~\cite{Berezhkovskii2013} and identifies the transition-state ensemble as states making up the $q_{\st}=0.5$ isocommittor surface~\cite{Berezhkovskii2005}. 

In this Letter, we derive a novel information-theoretic justification of the committor as the reaction coordinate. We show how selecting the transition-path ensemble (the set of trajectories from reactant to product) from \stkoutR{the equilibrium ensemble} {\MLR a long ergodic equilibrium trajectory} results in entropy production that precisely equals the information generated by system dynamics about the reactivity of trajectories. 

The components of entropy production and information generation due to an arbitrary system coordinate are also equal; this reveals equivalent thermodynamic and information-theoretic measures of the suitability of low-dimensional collective variables that encode information relevant for describing reaction mechanisms. The committor is a single coordinate that preserves all system entropy production and distills all system information about reactivity, giving further support for its role as the reaction coordinate.

\emph{Information-theoretic formulation of the committor as reaction coordinate.}---Consider a multidimensional system $\St$ evolving according to Markovian dynamics governed by the master equation~\cite{Zwanzig2001}, $\md_t p(\st) = \sum_{\st'} T_{\st \st'} p(\st')$, where $T_{\st \st'}$ is the transition rate from state $\st' \to \st$ and $p(\st)$ is the probability of state $\st$. We assume the transition rates obey detailed balance~\cite{Zwanzig2001} and the system is in equilibrium with its environment so that $p(\st)=\pi(\st)$, the equilibrium probability of $\st$. We study the transition-path ensemble (TPE), the set of trajectories that leave one subset of states $A \in \St$ and next visit a distinct subset $B \in \St \setminus A$ before $A$. In most applications, $A$ and $B$ are metastable states separated by a dynamical barrier; following Refs.~\cite{Metzner2009,Vanden-Eijnden2014}, we only assume that $A$ and $B$ do not overlap and lack direct transitions, i.e., $T_{\st \st'}=0$ for $\st' \in A$ and $\st \in B$.

\begin{figure}
    \centering
    \includegraphics[width=\linewidth]{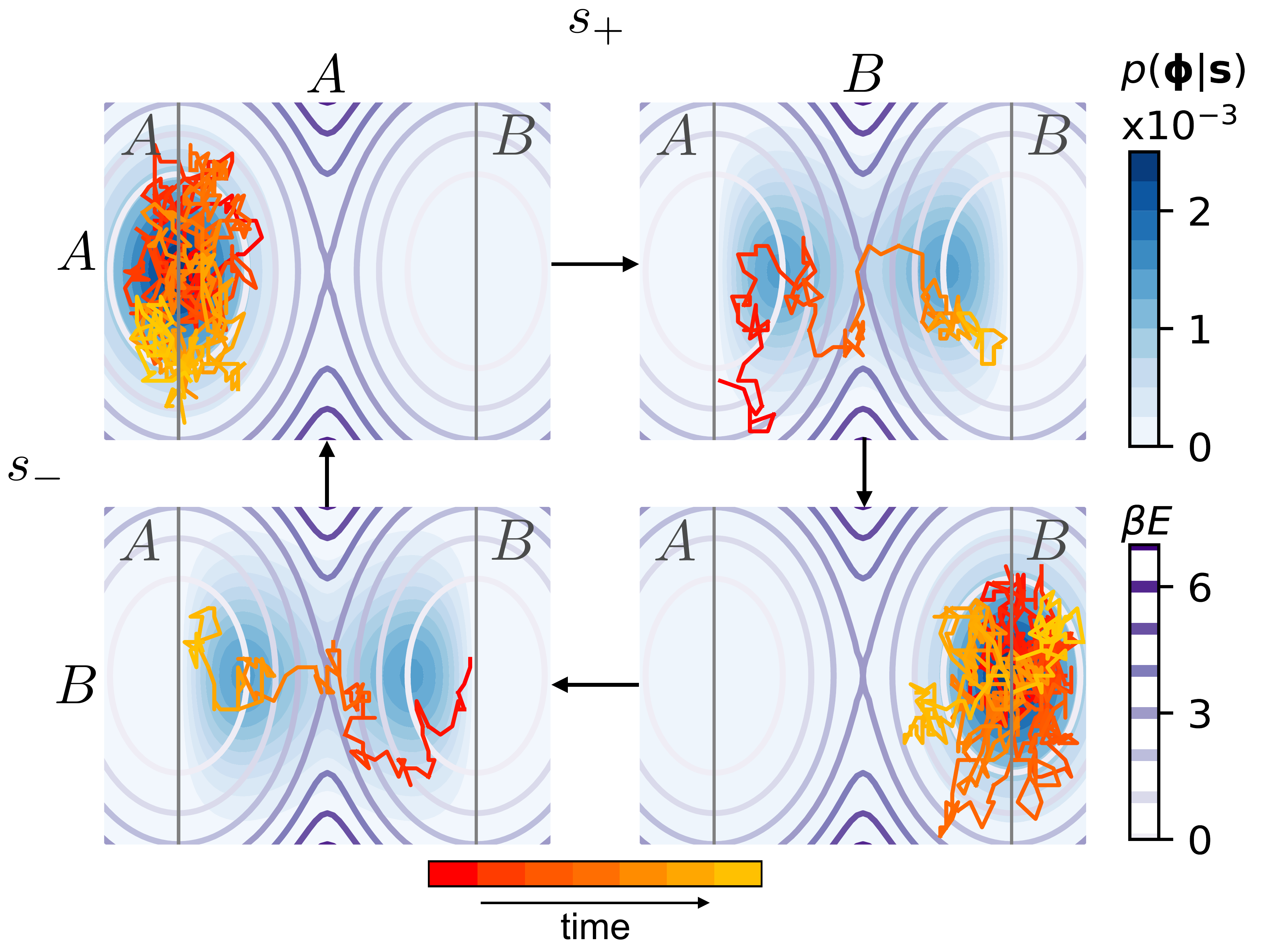}
    \caption{Partitioning 
    \stkoutR{an equilibrium trajectory}
    {\MLR a long ergodic equilibrium supertrajectory}
    into subensembles based on trajectory outcome $\Out$ and origin $\Org$. Contours: example double-well potential energy. Heat map: probability distribution $p(\st | \subens)$ of system state conditioned on trajectory subensemble $\subens$. Solid curves: representative trajectories from each subensemble. The forward (reverse) TPE in the top-right (bottom-left) panel has net flux of trajectories from $A \to B$ ($B \to A$). The top-left (bottom-right) panel shows the stationary subensemble from $A \to A$ ($B \to B$).}
    \label{fig:outcome_origin}
\end{figure}

The TPE can be formed by selecting from \stkoutR{the equilibrium ensemble} {\MLR a long ergodic equilibrium supertrajectory the} trajectory segments that leave $A$ and reach $B$ before $A$. Transition paths are therefore selected based on the trajectory outcome $\Out$ (the next mesostate ($A$ or $B$) visited by the system) and origin $\Org$ (the mesostate most recently visited by the system). {\MLR This partitions the supertrajectory into four trajectory subensembles, each with particular $\subens \equiv (\org,\out)$: The forward (reverse) transition-path ensemble is the set of trajectory segments with $\subens=(A,B)$ [$\subens=(B,A)$], and the stationary subensemble from $A \to A$ ($B \to B$) has \stkout{$\subens=(A,B)$} {\ML $\subens=(A,A)$} [$\subens=(B,B)$], as depicted in Fig.~\ref{fig:outcome_origin}. Every trajectory segment in the forward TPE has a corresponding equally probable time-reversed trajectory segment in the reverse TPE.} 

{\MLR At any time during the equilibrium supertrajectory, we define random variables $\St$ and $\Subens$, respectively, denoting the current system state and trajectory subensemble, with $p(\st,\subens)$ the joint distribution that the system is currently in state $\st$ and is currently on a trajectory segment with respective origin and outcome $\subens=\{\org,\out\}$. Since the system dynamics are Markovian, the trajectory outcome and origin are conditionally independent given current state $\st$, so the joint distribution can be factored as $p(\st,\subens)=\pi(\st)p(\out|\st)p(\org|\st)$~\cite{Metzner2009}. The conditional probabilities of trajectory outcome and origin given current state $\st$ are} \stkoutR{which have conditional probabilities}
\begin{subequations} \label{eq:committor_defn}
\begin{align}
    p(\Out=B\,|\,\st) &= q^+_{\st} \\
    p(\Org=A\,|\,\st) &= q^-_{\st}
    \>.
\end{align} 
\end{subequations}
Here $q^+_{\st}$ is the forward committor, the probability that the system currently in state $\st$ will next reach $B$ before $A$, and $q^-_{\st}$ is the backward committor, the probability that the system (currently in $\st$) was more recently in mesostate $A$ than in $B$. The committors obey boundary conditions $q^+_{\st} = 0$ and $q^-_{\st} = 1$ for $\st \in A$, and $q^+_{\st} = 1$ and $q^-_{\st} = 0$ for $\st \in B$. Since the system is in equilibrium and the transition rates obey detailed balance, $q^-_{\st} = 1-q^+_{\st}$~\cite{Metzner2009}, a single committor (without loss of generality, the forward committor $q^+_{\st}$) provides information about both the outcome and origin of the trajectory {\MLR segment}, so we refer to $q^+_{\st}$ as the reaction coordinate.

\stkoutR{The outcome $\Out$ and origin $\Org$ partition the equilibrium ensemble into four trajectory subensembles, each with particular $\subens \equiv (\org,\out)$: the forward (reverse) transition-path ensemble is the set of trajectories with $\subens=(A,B)$ ($\subens=(B,A)$), and the stationary subensemble from $A \to A$ ($B \to B$) has \stkout{$\subens=(A,B)$} {\ML $\subens=(A,A)$} ($\subens=(B,B)$), as depicted in Fig.~\ref{fig:outcome_origin}. Every trajectory in the forward TPE has a corresponding equally probable time-reversed trajectory in the reverse TPE.}

During \stkoutR{the system's equilibrium dynamics, it} {\MLR the equilibrium supertrajectory, the system} continually evolves from $A$ to $B$ and $B$ to $A$, completing a unidirectional cycle through each subensemble with stochastic transition times depending on underlying microscopic dynamics. Transition-path theory~\cite{Vanden-Eijnden2014,Berezhkovskii2019,E2006} derives quantitative properties (reaction rate and free-energy difference) of the $A\to B$ reaction from the equilibrium probability flux of subensemble transitions 
\begin{align}
\label{eq:nuS}
\nu_{\Subens}=\sum_{\st \notin A, \st' \in A} T_{\st \st'} \pi(\st') q^+_{\st}
\end{align}
and the respective marginal probabilities $p(\out)$ and $p(\org)$:
\begin{subequations} \label{eq:rxn_parameters}
	\begin{align}
		k_{AB} &= & &\frac{\nu_{\Subens}}{p(\Org=A)} & &= & &\frac{\nu_{\Subens}}{p(\Out=A)} \\
		\quad k_{BA} &= & &\frac{\nu_{\Subens}}{p(\Org=B)} & &= & &\frac{\nu_{\Subens}}{p(\Out=B)} \\
		\beta \Delta F_{AB} &= & \ln \, &\frac{p(\Org=A)}{p(\Org=B)} & &= & \ln \, &\frac{p(\Out=A)}{p(\Out=B)} \>,
	\end{align}
\end{subequations} 
{\ML where $k_{AB}$ ($k_{BA}$) is the rate constant for the $A \to B$ ($B \to A$) transition and $\Delta F_{AB} \equiv F_B-F_A$ is the free-energy difference.}
Mesoscopic reaction properties are therefore derived from information about the subensembles, specifically the proportion of time spent in each subensemble and how frequently the subensemble switches. 

The reaction coordinate should be maximally informative about the current subensemble. This is precisely quantified by mutual information, a nonlinear statistical measure of the relationship between two random variables, specifically quantifying the reduction of uncertainty {\ML (given by Shannon entropy $H(X) \equiv -\sum_x p(x) \ln p(x)$)} about one random variable from measuring another~\cite{Cover2006}:
\begin{equation} \label{eq:MI_state_subens}
    I(\Subens;\St) \equiv
    \sum_{\st,\subens} p(\st,\subens) \ln \frac{p(\st,\subens)}{\pi(\st) p(\subens)} \>,
\end{equation}
where \stkoutR{$p(\st,\subens)=\pi(\st)p(\out|\st)p(\org|\st)$ is the joint probability the system is in state $\st$ and on a trajectory with outcome $\out$ and origin $\org$,} {\ML \stkoutR{and} $p(\subens)=\sum_{\st} p(\st,\subens)$ is the marginal probability that the system is {\MLR currently} on a trajectory {\MLR segment} with outcome and origin $\subens=(\org,\out)$. Operationally, $p(\subens)$ can be estimated from the proportion of time $\tau_{\subens}$ spent in subensemble $\subens$ during \stkoutR{a} 
{\MLR an equilibrium super}trajectory of length $\tau$, $p(\subens) = \lim_{\tau \to \infty} \tau_{\subens}/\tau$}. If the committor depends only on a one-dimensional coordinate $\X \in \St$ (i.e. $q_{\st}=q_{\x}$), then $\X$ is a sufficient statistic for the mutual information between trajectory subensemble and full system state, i.e., $I(\Subens;\St)=I(\Subens;\X)$. In this sense, the committor is the ``optimal'' reaction coordinate, since it is maximally informative about the trajectory subensemble given a measurement of system state. This is our first major result.

Physically, the trajectory outcome and origin (and hence the committors) represent uncertainty in the state of the environment. Classical mechanics assumes a constant-energy universe (system {\ML $\St$} plus environment {\ML $\Env$}) governed by deterministic dynamics so that the outcome and origin of the trajectory initiated from a given state of system and environment are deterministic (and can be determined by integrating the state of the universe forward and backward in time until the system reaches $A$ or $B$){\ML , i.e., $p(\subens|\st,\env)$ is either 0 or 1. This partitions the state space of the universe into four quadrants corresponding to each trajectory subensemble, with each state $(\st,\env)$ belonging to only one subensemble; thus the uncertainty about the trajectory subensemble given a state of the universe is zero, $H(\Subens|\St,\Env)\equiv-\sum_{\st,\env,\subens} p(\st,\env,\subens) \ln p(\subens|\st,\env) = 0$. In this case, the mutual information between the universe and trajectory subensemble is the uncertainty about the trajectory subensemble, $I(\Subens;\St,\Env)=H(\Subens)-H(\Subens|\St,\Env)=H(\Subens)$; the measurement of the state of the universe fully determines the trajectory outcome and origin}. 

However, we typically do not resolve the microstate of the environment, instead coarse-graining its interaction with the system into friction and fluctuations~\cite{Zwanzig2001}. \stkout{When only the system state is resolved, the outcome and origin} 
{\ML Measurement of the system state alone does not fully determine the trajectory outcome and origin, which} 
become random variables with positive 
{\ML conditional}
Shannon entropy $H(\Subens| \St)\equiv-\sum_{\subens,\st} p(\st,\subens) \ln p(\subens | \st)>0$ reflecting uncertainty \stkout{about the environment} {\ML in the state of the environment that is relevant to classification of the current subensemble}.

\emph{Transition-path thermodynamics.}---The joint dynamics of $(\St,\Subens)$ is given by the master equation
\begin{equation} \label{eq:joint_master_eqn}
    \md_t p(\st,\subens) = \sum_{\st',\subens'} T^{\subens \subens'}_{\st \st'} \, p(\st',\subens')
\end{equation}
where the $(\st',\subens') \to (\st,\subens)$ transition rate is (see Supplemental Material~\ref{appendix_transition_rates}~\cite{SM})
\begin{equation} \label{eq:TPE_transition_rates}
T^{\subens \subens'}_{\st \st'} =
\begin{cases} 
    T^{\subens}_{\st \st'} \equiv T_{\st \st'} \frac{p(\out| \st)}{p(\out| \st')}, & \subens' = \subens \\
    T_{\st \st'} p(\Out=B|\st) \, , &
    \begin{cases}
        \st' \in A \> {\rm{, }} \> \st \notin A \> \rm{, } \\
        \subens'=(A,A) \> {\rm{, }} \> \subens=(A,B) \\
    \end{cases} \\
    \vspace{-2.5ex}\\
    T_{\st \st'} p(\Out=A|\st) \, , &
    \begin{cases}
        \st' \in B \> {\rm{, }} \> \st \notin B \> \rm{, } \\
        \subens'=(B,B) \> {\rm{, }} \> \subens=(B,A)  \\
    \end{cases} \\
    \vspace{-2.5ex}\\
    T_{\st \st'}/p(\Out=A|\st') \, , &
    \begin{cases}
        \st' \notin A \> {\rm{, }} \> \st \in A \> \rm{, } \\
        \subens'=(B,A) \> {\rm{, }} \> \subens=(A,A)  \\
    \end{cases} \\
    \vspace{-2.5ex}\\
    T_{\st \st'}/p(\Out=B|\st') \, , &
    \begin{cases}
        \st' \notin B \> {\rm{, }} \> \st \in B \> \rm{, } \\
        \subens'=(A,B) \> {\rm{, }} \> \subens=(B,B)  \\
    \end{cases} \\
    -\sum
    \limits_{\substack{\st'' \neq \st'\\ \subens'' \neq \subens'}} 
    \, T_{\st'' \st'}^{\subens'' \subens'}, & \st = \st' \> {\rm{,}} \> \subens = \subens' \\
    0 & \rm{otherwise} 
\end{cases}
\>.
\end{equation}
The top transition does not change the subensemble{\ML, and biases transitions within subensemble $\subens$ toward states with higher probability of trajectory outcome $\out$}. The middle four transitions switch subensembles and are unidirectional, {\ML contributing to the} \stkout{with} probability flux $\nu_{\Subens}$ [Eq.~\eqref{eq:nuS} and Supplemental Material Eq.~\eqref{eq:uni_prob_flux}]. {\ML These joint dynamics are Markovian: Since the underlying system dynamics are Markovian, the transition rates~\eqref{eq:TPE_transition_rates} do not depend on the trajectory origin $\org$, and the outcome $\out$ does not induce dependence on earlier system states.}
Considered alone, system dynamics are at equilibrium and microscopically reversible; adding the trajectory outcome and origin variables (that are not functions of system state and explicitly depend on the past and future) breaks time-reversal symmetry, producing \stkout{the} absolutely irreversible trajectory-subensemble transitions
{\ML and time-asymmetric system transitions within a given subensemble}. 

{\ML To quantify the time asymmetry for a particular $\st' \to \st$ transition in subensemble $\subens$, we combine \eqref{eq:TPE_transition_rates} Bayes' rule, and the equilibrium detailed-balance relation $T_{\st \st'} \pi(\st')=T_{\st' \st} \pi(\st)$ to derive a local detailed-balance relation,
\begin{equation} \label{eq:LDB_relation}
    \frac{T^{\subens}_{\st \st'}p(\st'|\subens)}{T^{\subens}_{\st' \st}p(\st|\subens)} = \frac{p(\org|\st') p(\out|\st)}{p(\org|\st) p(\out|\st')}\ . 
\end{equation} 
The $A \to A$ (and analogously $B \to B$) stationary subensemble has $\out=\org=A$, and due to system detailed balance $p(\Org = A|\st) = p(\Out = A|\st)$, so the rhs is unity and detailed balance holds for transitions within stationary subensembles. The reactive subensembles (forward or reverse TPE) have different trajectory outcome and origin so the rhs side differs from unity, leading to a detailed-balance-breaking flux (and hence entropy production) along particular transitions within these subensembles.}

\stkout{For}
{\ML Within}
a fixed subensemble $\subens$, the \stkout{joint dynamics produce} net trajectory flux {\ML is}
\begin{subequations}
\begin{align} \label{eq:TPE_net_flux}
	J^{\subens}_{\st \st'} &= T^{\subens}_{\st \st'} p(\st', \subens) - T^{\subens}_{\st' \st} p(\st, \subens) \\
	&= \big[ p(\org| \st') p(\out| \st)-p(\org| \st) p(\out| \st') \big] T_{\st \st'} \pi(\st') \> .
\end{align}
\end{subequations}
The second equality follows from $p(\st,\subens)=p(\subens|\st)\pi(\st)$; the conditional independence of $\out$ and $\org$ given state $\st$, i.e., $p(\out,\org|\st)=p(\out|\st)p(\org|\st)$; and substitution for $T^{\subens}_{\st' \st}$ using \eqref{eq:TPE_transition_rates}. The stationary subensembles ($A \to A$ and $B \to B$) have no net flux because each trajectory {\MLR segment} and its time-reversed counterpart occur at equal rates within the same subensemble. In contrast, the forward and reverse TPEs have net trajectory flux since each transition path and its time-reversed counterpart occur in different subensembles.
\stkout{This suggests that the joint dynamics of $(\St,\Subens)$ have irreversible positive entropy production.}
{\ML (Our procedure of effectively replicating the state space and introducing opposing fluxes in the replicas by modification of the transition rates bears similarity to nonreversible Markov chains obeying skew detailed balance used to speed convergence to a stationary distribution~\cite{Turitsyn2011}.)}

We decompose (see Supplemental Material~\ref{appendix_entropy_transitions}~\cite{SM}) the change in joint entropy $H(\St,\Subens)\equiv-\sum_{\st,\subens} p(\st,\subens) \ln p(\st,\subens)$ at steady state~\cite{Busiello2020,Esposito2012} into
\begin{subequations}
\begin{align} 
    0 &= \md_t H(\St,\Subens) \label{eq:joint_entropy_production} \\
    &= \underbrace{\sum_{\subens} p(\subens) \sum_{\st,\st'} T^{\subens}_{\st \st'} p(\st'|\subens) \ln \frac{T^{\subens}_{\st \st'} p(\st'|\subens)}{T^{\subens}_{\st' \st} p(\st|\subens)}}_{\langle \dot{\Sigma} \rangle} \nonumber \\
    & \quad - 2 \ \underbrace{\sum_{\st,\st', \out} T^{\out}_{\st \st'} p(\st',\out) \ln \frac{p(\out|\st)}{p(\out|\st')}}_{\dot{I}^{\St}(\Out; \St)} \label{eq:EPR_Idot} \>,
\end{align}
\end{subequations}
where $\langle \dot{\Sigma} \rangle = \sum_{\subens} p(\subens) \dot{\Sigma}_{\subens}$ is the subensemble-weighted average of the irreversible entropy production rate $\dot{\Sigma}_{\subens}$ conditioned on subensemble $\subens$, which quantifies the time irreversibility of system dynamics \stkout{in} {\ML within} that subensemble. 
$\dot{I}^{\St}(\Out; \St) \ge 0$ is the rate of change in mutual information between the trajectory outcome and system state due to system dynamics in a fixed subensemble~\cite{Horowitz2014}. Rearranging Eq.~\eqref{eq:EPR_Idot} gives (see Supplemental Material~\ref{appendix_MI_entropy}~\cite{SM}): 
\begin{subequations} \label{eq:irreversible_entropy}
\begin{align}
0 \le \langle \dot{\Sigma} \rangle &= 2 \dot{I}^{\St}(\Out; \St) \label{eq:entropy_info_outcome} \\
&= \dot{I}^{\St}(\Out; \St) - \dot{I}^{\St}(\Org; \St) \>, \label{eq:entropy_info_difference}
\end{align}
\end{subequations}
where $\dot{I}^{\St}(\Org;\St) \le 0$ is the rate of change in mutual information between trajectory origin and system state due to $\St$ dynamics in a fixed subensemble. 
\stkout{As the system evolves, the information the system state carries about the trajectory outcome increases, while the information the system state carries about the origin decreases.}
{\ML These information rates reflect the dependence of the trajectory outcome and origin variables on the past and future states of the system: As the system evolves, uncertainty about the outcome $\Out$ diminishes, increasing the information the current system state carries about $\Out$, while uncertainty (given current system state $\St$) about the origin $\Org$ increases, decreasing information $\St$ carries about the $\Org$.}

Since the stationary subensembles have no entropy production ($\dot{\Sigma}_{\subens=(A,A)} =\dot{\Sigma}_{\subens=(B,B)}=0$), Eq.~\eqref{eq:entropy_info_outcome} reduces to an equation for a single subensemble,
\begin{equation} 
\label{eq:info_equals_TPE_entropy}
    0 \le p_{\TPE} \dot{\Sigma}_{\TPE} = \dot{I}^{\St}(\Out;\St) \>.
\end{equation} 
This equates the rate $\dot{I}^{\St}(\Out;\St)$ of generating information about the outcome with the product of the entropy production rate of a reactive subensemble \stkout{(forward or reverse TPE)} $\dot{\Sigma}_{\TPE} =\dot{\Sigma}_{\subens=(A,B)} =\dot{\Sigma}_{\subens=(B,A)}$ and that subensemble's marginal probability $p_{\TPE}=p(\Subens=(A,B))=p(\Subens=(B,A))$.
{\MLR Although the supertrajectory is at equilibrium with no entropy production, $\dot{\Sigma}_{\TPE}$ physically represents the dissipation that would be necessary in a system evolving according to the TPE's detailed-balance-breaking transition rates (top line of (6) for $\org \neq \out$).}
Equation~\eqref{eq:info_equals_TPE_entropy} is our second major result: The entropy production in a reactive subensemble equals the information generated about the reactivity of trajectories. 

When the state space $\St$ is continuous, we derive (see Supplemental Material~\ref{appendix_TD_metric}~\cite{SM}) a Fisher-information metric $\mathcal{I}(\st)$ that imposes an information geometry on the state space~\cite{Amari2016,Nielsen2020}. The metric measures distance on the reaction coordinate (committor) as the system evolves and thereby defines a reaction-coordinate length $\mathcal{L}_{AB}$. From this, the TPE entropy production is
\begin{equation} \label{eq:TD_length}
    \dot{\Sigma}_{\TPE} \approx \frac{\mathcal{L}^2_{AB}}{2 \tau_{\TPE}} \>,
\end{equation}
where $\tau_{\TPE}$ is the mean duration of a transition path. This relates the TPE entropy production to the squared length between $A$ and $B$ along the reaction coordinate.

\emph{Bipartite dynamics.}---We now demonstrate how the TPE entropy production quantitatively measures the relevance of an arbitrary coordinate to the reaction. We assume bipartite dynamics~\cite{Hartich2014,Barato2013}, essentially that instantaneous transitions only happen in either a one-dimensional coordinate $\X$ or in all other degrees of freedom $\Y$ making up the system state $\St=(\X,\Y)$: 
\begin{equation} 
T_{\x \x',\y \y'} = 
\begin{cases} 
    T_{\x \x',\y} &  \x \neq \x' \> {\rm{,}} \> \y = \y' \\
    T_{\x,\y \y'} &  \x = \x' \> {\rm{,}} \> \y \neq \y' \\
    -\sum
    \limits_{\substack{\x'' \neq \x'\\ \y'' \neq \y'}} 
    \, T_{\x'' \x',\y'' \y'} & \x = \x' \> {\rm{,}} \> \y = \y' \\ 
    0 & \rm{otherwise} 
\end{cases} \>.
\end{equation} 
{\ML Dynamics that do not obey the bipartite assumption introduce further complications in unambiguously partitioning the entropy production between coordinates~\cite{Chetrite2019}.} 

Combining Eqs.~\eqref{eq:committor_defn}, \eqref{eq:TPE_net_flux}, and \eqref{eq:EPR_Idot} gives the full TPE entropy production as a function of the forward committor,
\begin{equation}
     p_{\TPE} \dot{\Sigma}_{\TPE} = \tfrac{1}{2} \sum_{\st,\st'} T_{\st \st'} \pi(\st') (q^+_{\st} - q^+_{\st'}) \ln \frac{q^+_{\st} (1-q^+_{\st'})}{q^+_{\st'} (1-q^+_{\st})} \>,
\end{equation}
which splits into contributions from the two transition types:
\begin{subequations} \label{eq:bipartite_dissipation}
\begin{align}
    &p_{\TPE} \dot{\Sigma}_{\TPE} = p_{\TPE} \dot{\Sigma}_{\TPE}^{\X} + p_{\TPE} \dot{\Sigma}_{\TPE}^{\Y} \\
    &= \tfrac{1}{2} \sum_{\x,\x',\y} T_{\x \x',\y} \pi(\x',\y) (q^+_{\x \y}- q^+_{\x' \y}) \ln{\frac{q^+_{\x' \y} (1-q^+_{\x \y})}{q^+_{\x \y} (1-q^+_{\x' \y})}} \label{eq:bipartite_dissipationXY} \\
    &\quad + \tfrac{1}{2} \sum_{\x,\y,\y'} T_{\x,\y \y'} \pi(\x,\y') (q^+_{\x \y} - q^+_{\x \y'}) \ln{\frac{q^+_{\x \y} (1-q^+_{\x \y'})}{q^+_{\x \y'} (1-q^+_{\x \y})}} \nonumber \> .
\end{align}
\end{subequations}

The same decomposition holds for the information rate~\cite{Horowitz2014}, so that TPE entropy production due to $\X$ dynamics is equal to the information rate (due to $X$ dynamics) between $\St$ and $\Out$:
\begin{equation}
     p_{\TPE} \dot{\Sigma}_{\TPE}^{\X} = \dot{I}^{\X}(\Out; \St) \>.
\end{equation}
This is our third major result: The entropy production due to dynamics of coordinate $X$ equals the mutual information generated by $\X$ dynamics, thereby quantifying the relevance of $X$ transitions to identifying the current subensemble and highlighting those transitions that are ``correlated'' with reactive trajectories and therefore important to the reaction mechanism.

In particular, for $\X^*$ determining the committor and $\Y^*$ orthogonal degrees of freedom that are therefore not relevant to the reaction ($q_{\x \y}=q_{\x}$), the entropy production rate due to $\Y^*$ dynamics is [simplifying Eq.~\eqref{eq:bipartite_dissipationXY}]:
\begin{subequations}
\begin{align}
    \dot{\Sigma}_{\TPE}^{\Y^*} &= \sum_{\x,\y,\y'} T_{\x, \y \y'} \pi(\x, \y') (q^+_{\x} -q^+_{\x}) \ln \frac{q^+_{\x} (1-q^+_{\x})}{q^+_{\x} (1-q^+_{\x})} \\
    &= 0 \>.
\end{align}
\end{subequations}
Therefore $\dot{\Sigma}_{\TPE}^{\X^*} = \dot{\Sigma}_{\TPE}$. This is additional confirmation that the committor is the reaction coordinate, in that it provides a thermodynamically complete coarse-grained representation of the transition-path ensemble, fully accounting for its entropy production~\cite{Esposito2012}. 

We illustrate with overdamped dynamics in a double-well energy landscape [Fig.~\ref{fig:bistable_figure}(a), details in Supplemental Material~\ref{appendix_bistable}~\cite{SM}]. To exemplify the typical situation where the reaction coordinate is not known \textit{a priori} and coordinates are thus chosen based on convenience or intuition, fixed system coordinates $(x,y)$ lie at an angle $\theta$ to the correct reaction coordinate, the linear coordinate passing through both energy minima. For $\theta=0^{\circ}$, $X$ is the reaction coordinate, $Y$ is an orthogonal bath mode~\cite{Li2016}, and $X$ dynamics fully capture the TPE entropy production without $Y$ contribution. Figure~\ref{fig:bistable_figure}(b) shows that as the underlying energy landscape is rotated relative to system coordinates, the $X$-coordinate entropy production decreases and $Y$-coordinate entropy production increases, with equal contribution at $\theta=45^{\circ}$. The entropy production for each coordinate is proportional to the squared Euclidean distance between $A$ and $B$ projected onto each coordinate, $\dot{\Sigma}_{\TPE}^{X} (\theta) \propto \cos^2 \theta$ and $\dot{\Sigma}_{\TPE}^{Y} (\theta) \propto \sin^2 \theta$.

\begin{figure}
    \centering
    \includegraphics[width=\linewidth]{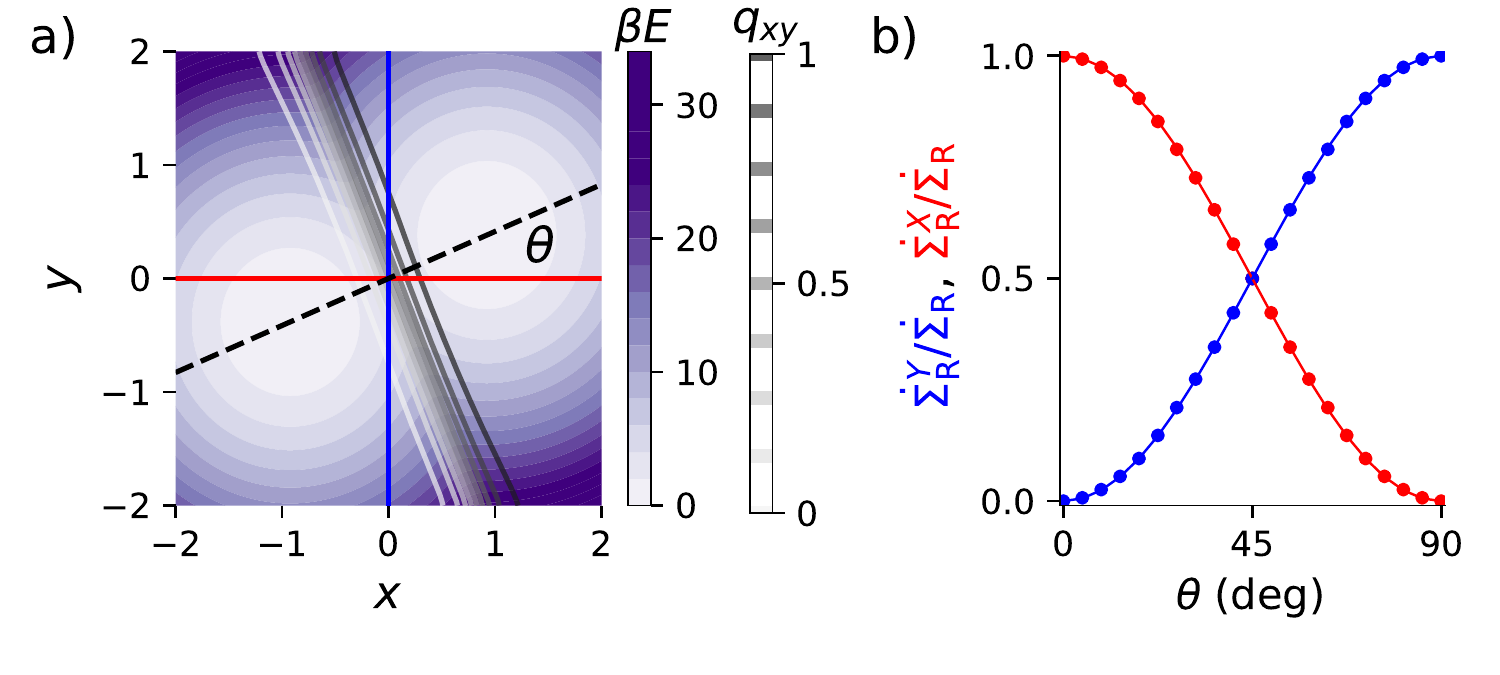} 
    \caption{a) Double-well energy landscape (purple) with fixed $x$ axis (red solid) and $y$ axis (blue solid), and rotated reaction coordinate (black dashed). Gray contours: forward committor. b) Share of TPE entropy production due to coordinate $X$ (red dots) and $Y$ (blue dots), as a function of reaction-coordinate angle $\theta$. Red solid: $\cos^2 \theta$. Blue solid: $\sin^2 \theta$.}
    \label{fig:bistable_figure}
\end{figure}

\emph{Discussion.}---We have derived the information thermodynamics of a system undergoing reactions between distinct state-space subsets $A$ and $B$, making a fundamental connection between transition-path theory, information theory, and stochastic thermodynamics. Partitioning \stkoutR{the equilibrium ensemble} {\MLR a long ergodic equilibrium trajectory} into reactive and nonreactive subensembles results in entropy production for system dynamics in the reactive subensembles {\MLR (physically representing the dissipation needed to implement the detailed-balance-breaking transition rates of the TPE)}, which in turn identifies transitions that are relevant to the overall reaction mechanism. This rigorous equality between TPE entropy production and informativeness of dynamics also holds for an arbitrary coordinate, revealing parallel stochastic-thermodynamic and information-theoretic measures of the relevance of collective variables to the system reaction, that are each maximized by the committor. 

This work has implications for the identification of important collective variables and analysis of reaction mechanisms. While the committor provides a microscopically detailed reaction coordinate that maps each system microstate to a scalar value, it does not immediately identify physically meaningful collective variables (e.g., internal molecular coordinates) that are relevant to the reaction~\cite{Bolhuis2015,Peters2016,Johnson2012}. Our results have indicated that relevant coordinates are identified by entropy production in the transition-path ensemble;
\stkoutR{one could therefore find the most dissipative modes from principal-component analysis~\cite{Gnesotto2020} or estimate the entropy production for motion in each coordinate~\cite{Seifert2019} from a sample of transition paths to rank a reduced set of coordinates by their contribution to the reaction. Our results show that }
{\MLR thus}
partitioning the entropy production between multiple relevant collective variables for which one has physical intuition can provide a low-dimensional model that allows increased insight into the reaction mechanism.

{\MLR More concretely, this connection we have established between transition-path theory and stochastic thermodynamics suggests a novel method for rigorously grounded inference of reaction coordinates: generate an ensemble of transition paths using transition-path sampling~\cite{Dellago1998,Bolhuis2002} or related algorithms~\cite{vanErp2003,Allen2005,Elber2004}; estimate entropy production along chosen coordinates~\cite{Seifert2019,Li2019a,Skinner2021} or identify linear combinations of coordinates producing the most entropy using dissipative components analysis~\cite{Gnesotto2020}; use these most dissipative coordinates to enhance sampling of transition paths; and through further iteration identify system coordinates producing the most entropy in the transition-path ensemble and hence of most relevance to the reaction.}

Machine-learning approaches to solve for high-dimensional committor coordinates~\cite{Khoo2019,Li2019,Rotskoff2021} or find low-dimensional reaction models that retain predictive power~\cite{Ma2005,Wang2021} are active areas of research~\cite{Wang2020}. The information-theoretic and thermodynamic perspectives on reactive trajectories described in this Letter provide guidance to the development of data-intensive automated methods to infer these models and their corresponding reaction mechanisms.

\vspace{2ex}
This work was supported by Natural Sciences and Engineering Research Council of Canada (NSERC) Canada Graduate Scholarships Masters and Doctoral (MDL), an NSERC Discovery Grant (DAS), and a Tier-II Canada Research Chair (DAS). The authors thank Jannik Ehrich (SFU Physics) for enlightening feedback on the manuscript.


\begin{thebibliography}{48}
\expandafter\ifx\csname natexlab\endcsname\relax\def\natexlab#1{#1}\fi
\expandafter\ifx\csname bibnamefont\endcsname\relax
  \def\bibnamefont#1{#1}\fi
\expandafter\ifx\csname bibfnamefont\endcsname\relax
  \def\bibfnamefont#1{#1}\fi
\expandafter\ifx\csname citenamefont\endcsname\relax
  \def\citenamefont#1{#1}\fi
\expandafter\ifx\csname url\endcsname\relax
  \def\url#1{\texttt{#1}}\fi
\expandafter\ifx\csname urlprefix\endcsname\relax\def\urlprefix{URL }\fi
\providecommand{\bibinfo}[2]{#2}
\providecommand{\eprint}[2][]{\url{#2}}

\bibitem[{\citenamefont{Eyring}(1935)}]{Eyring1935}
\bibinfo{author}{\bibfnamefont{H.}~\bibnamefont{Eyring}}, \bibinfo{journal}{J.
  Chem. Phys.} \textbf{\bibinfo{volume}{3}}, \bibinfo{pages}{107}
  (\bibinfo{year}{1935}).

\bibitem[{\citenamefont{Evans and Polanyi}(1935)}]{Evans1935}
\bibinfo{author}{\bibfnamefont{M.~G.} \bibnamefont{Evans}} \bibnamefont{and}
  \bibinfo{author}{\bibfnamefont{M.}~\bibnamefont{Polanyi}},
  \bibinfo{journal}{Trans. Faraday Soc.} \textbf{\bibinfo{volume}{31}},
  \bibinfo{pages}{875} (\bibinfo{year}{1935}).

\bibitem[{\citenamefont{Kramers}(1940)}]{Kramers1940}
\bibinfo{author}{\bibfnamefont{H.~A.} \bibnamefont{Kramers}},
  \bibinfo{journal}{Physica} \textbf{\bibinfo{volume}{7}}, \bibinfo{pages}{284}
  (\bibinfo{year}{1940}).

\bibitem[{\citenamefont{Peters}(2016)}]{Peters2016}
\bibinfo{author}{\bibfnamefont{B.}~\bibnamefont{Peters}},
  \bibinfo{journal}{Annu. Rev. Phys. Chem} \textbf{\bibinfo{volume}{67}},
  \bibinfo{pages}{669} (\bibinfo{year}{2016}).

\bibitem[{\citenamefont{Bolhuis and Dellago}(2015)}]{Bolhuis2015}
\bibinfo{author}{\bibfnamefont{P.~G.} \bibnamefont{Bolhuis}} \bibnamefont{and}
  \bibinfo{author}{\bibfnamefont{C.}~\bibnamefont{Dellago}},
  \bibinfo{journal}{Eur. Phys. J. Spec. Top.} \textbf{\bibinfo{volume}{224}},
  \bibinfo{pages}{2409} (\bibinfo{year}{2015}).

\bibitem[{\citenamefont{Dellago et~al.}(1998)\citenamefont{Dellago, Bolhuis,
  Csajka, and Chandler}}]{Dellago1998}
\bibinfo{author}{\bibfnamefont{C.}~\bibnamefont{Dellago}},
  \bibinfo{author}{\bibfnamefont{P.~G.} \bibnamefont{Bolhuis}},
  \bibinfo{author}{\bibfnamefont{F.~S.} \bibnamefont{Csajka}},
  \bibnamefont{and} \bibinfo{author}{\bibfnamefont{D.}~\bibnamefont{Chandler}},
  \bibinfo{journal}{J. Chem. Phys.} \textbf{\bibinfo{volume}{108}},
  \bibinfo{pages}{1964} (\bibinfo{year}{1998}).

\bibitem[{\citenamefont{E and Vanden-Eijnden}(2006)}]{E2006}
\bibinfo{author}{\bibfnamefont{W.}~\bibnamefont{E}} \bibnamefont{and}
  \bibinfo{author}{\bibfnamefont{E.}~\bibnamefont{Vanden-Eijnden}},
  \bibinfo{journal}{J. Stat. Phys.} \textbf{\bibinfo{volume}{123}},
  \bibinfo{pages}{503} (\bibinfo{year}{2006}).

\bibitem[{\citenamefont{E and Vanden-Eijnden}(2010)}]{E2010}
\bibinfo{author}{\bibfnamefont{W.}~\bibnamefont{E}} \bibnamefont{and}
  \bibinfo{author}{\bibfnamefont{E.}~\bibnamefont{Vanden-Eijnden}},
  \bibinfo{journal}{Annu. Rev. Phys. Chem} \textbf{\bibinfo{volume}{61}},
  \bibinfo{pages}{391} (\bibinfo{year}{2010}).

\bibitem[{\citenamefont{Li and Ma}(2014)}]{Li2014}
\bibinfo{author}{\bibfnamefont{W.}~\bibnamefont{Li}} \bibnamefont{and}
  \bibinfo{author}{\bibfnamefont{A.}~\bibnamefont{Ma}}, \bibinfo{journal}{Mol
  Simul.} \textbf{\bibinfo{volume}{40}}, \bibinfo{pages}{784}
  (\bibinfo{year}{2014}).

\bibitem[{\citenamefont{Peters et~al.}(2013)\citenamefont{Peters, Bolhuis,
  Mullen, and Shea}}]{Peters2013}
\bibinfo{author}{\bibfnamefont{B.}~\bibnamefont{Peters}},
  \bibinfo{author}{\bibfnamefont{P.~G.} \bibnamefont{Bolhuis}},
  \bibinfo{author}{\bibfnamefont{R.~G.} \bibnamefont{Mullen}},
  \bibnamefont{and} \bibinfo{author}{\bibfnamefont{J.-E.} \bibnamefont{Shea}},
  \bibinfo{journal}{J. Chem. Phys.} \textbf{\bibinfo{volume}{138}},
  \bibinfo{pages}{054106} (\bibinfo{year}{2013}).

\bibitem[{\citenamefont{Banushkina and Krivov}(2016)}]{Banushkina2016}
\bibinfo{author}{\bibfnamefont{P.~V.} \bibnamefont{Banushkina}}
  \bibnamefont{and} \bibinfo{author}{\bibfnamefont{S.~V.}
  \bibnamefont{Krivov}}, \bibinfo{journal}{WIREs Comput Mol Sci}
  \textbf{\bibinfo{volume}{6}}, \bibinfo{pages}{748} (\bibinfo{year}{2016}).

\bibitem[{\citenamefont{Berezhkovskii and Szabo}(2013)}]{Berezhkovskii2013}
\bibinfo{author}{\bibfnamefont{A.~M.} \bibnamefont{Berezhkovskii}}
  \bibnamefont{and} \bibinfo{author}{\bibfnamefont{A.}~\bibnamefont{Szabo}},
  \bibinfo{journal}{J. Phys. Chem. B} \textbf{\bibinfo{volume}{117}},
  \bibinfo{pages}{13115} (\bibinfo{year}{2013}).

\bibitem[{\citenamefont{Berezhkovskii and Szabo}(2005)}]{Berezhkovskii2005}
\bibinfo{author}{\bibfnamefont{A.}~\bibnamefont{Berezhkovskii}}
  \bibnamefont{and} \bibinfo{author}{\bibfnamefont{A.}~\bibnamefont{Szabo}},
  \bibinfo{journal}{J. Chem. Phys.} \textbf{\bibinfo{volume}{122}},
  \bibinfo{pages}{014503} (\bibinfo{year}{2005}).

\bibitem[{\citenamefont{Zwanzig}(2001)}]{Zwanzig2001}
\bibinfo{author}{\bibfnamefont{R.}~\bibnamefont{Zwanzig}},
  \emph{\bibinfo{title}{{Nonequilibrium statistical mechanics}}}
  (\bibinfo{publisher}{Oxford University Press}, \bibinfo{address}{New York},
  \bibinfo{year}{2001}).

\bibitem[{\citenamefont{Metzner et~al.}(2009)\citenamefont{Metzner, Schutte,
  and Vanden-Eijnden}}]{Metzner2009}
\bibinfo{author}{\bibfnamefont{P.}~\bibnamefont{Metzner}},
  \bibinfo{author}{\bibfnamefont{C.}~\bibnamefont{Schutte}}, \bibnamefont{and}
  \bibinfo{author}{\bibfnamefont{E.}~\bibnamefont{Vanden-Eijnden}},
  \bibinfo{journal}{SIAM Multiscale Model. Simul.}
  \textbf{\bibinfo{volume}{7}}, \bibinfo{pages}{1192} (\bibinfo{year}{2009}).

\bibitem[{\citenamefont{Vanden-Eijnden}(2014)}]{Vanden-Eijnden2014}
\bibinfo{author}{\bibfnamefont{E.}~\bibnamefont{Vanden-Eijnden}}, in
  \emph{\bibinfo{booktitle}{An introduction to Markov state models their
  application to long timescale molecular simulation}}, edited by
  \bibinfo{editor}{\bibfnamefont{G.~R.} \bibnamefont{Bowman}},
  \bibinfo{editor}{\bibfnamefont{V.~S.} \bibnamefont{Pande}}, \bibnamefont{and}
  \bibinfo{editor}{\bibfnamefont{F.}~\bibnamefont{Noe}}
  (\bibinfo{publisher}{Springer}, \bibinfo{year}{2014}),
  chap.~\bibinfo{chapter}{7}, pp. \bibinfo{pages}{91--100}.

\bibitem[{\citenamefont{Berezhkovskii and Szabo}(2019)}]{Berezhkovskii2019}
\bibinfo{author}{\bibfnamefont{A.~M.} \bibnamefont{Berezhkovskii}}
  \bibnamefont{and} \bibinfo{author}{\bibfnamefont{A.}~\bibnamefont{Szabo}},
  \bibinfo{journal}{J. Chem. Phys} \textbf{\bibinfo{volume}{150}},
  \bibinfo{pages}{054106} (\bibinfo{year}{2019}).

\bibitem[{\citenamefont{Cover and Thomas}(2006)}]{Cover2006}
\bibinfo{author}{\bibfnamefont{T.~M.} \bibnamefont{Cover}} \bibnamefont{and}
  \bibinfo{author}{\bibfnamefont{J.~A.} \bibnamefont{Thomas}},
  \emph{\bibinfo{title}{{Elements of information theory}}}
  (\bibinfo{publisher}{John Wiley {\&} Sons, Inc.}, \bibinfo{address}{Hoboken,
  New Jersey}, \bibinfo{year}{2006}), \bibinfo{edition}{2nd} ed.

\bibitem[{SM()}]{SM}
\bibinfo{note}{See Supplemental Material at [URL will be inserted by
  publisher].}

\bibitem[{\citenamefont{Turitsyn et~al.}(2011)\citenamefont{Turitsyn, Chertkov,
  and Vucelja}}]{Turitsyn2011}
\bibinfo{author}{\bibfnamefont{K.~S.} \bibnamefont{Turitsyn}},
  \bibinfo{author}{\bibfnamefont{M.}~\bibnamefont{Chertkov}}, \bibnamefont{and}
  \bibinfo{author}{\bibfnamefont{M.}~\bibnamefont{Vucelja}},
  \bibinfo{journal}{Physica D} \textbf{\bibinfo{volume}{240}},
  \bibinfo{pages}{410} (\bibinfo{year}{2011}).

\bibitem[{\citenamefont{Busiello et~al.}(2020)\citenamefont{Busiello, Gupta,
  and Maritan}}]{Busiello2020}
\bibinfo{author}{\bibfnamefont{D.~M.} \bibnamefont{Busiello}},
  \bibinfo{author}{\bibfnamefont{D.}~\bibnamefont{Gupta}}, \bibnamefont{and}
  \bibinfo{author}{\bibfnamefont{A.}~\bibnamefont{Maritan}},
  \bibinfo{journal}{Phys. Rev. Res.} \textbf{\bibinfo{volume}{2}},
  \bibinfo{pages}{1} (\bibinfo{year}{2020}).

\bibitem[{\citenamefont{Esposito}(2012)}]{Esposito2012}
\bibinfo{author}{\bibfnamefont{M.}~\bibnamefont{Esposito}},
  \bibinfo{journal}{Phys. Rev. E} \textbf{\bibinfo{volume}{85}},
  \bibinfo{pages}{041125} (\bibinfo{year}{2012}).

\bibitem[{\citenamefont{Horowitz and Esposito}(2014)}]{Horowitz2014}
\bibinfo{author}{\bibfnamefont{J.~M.} \bibnamefont{Horowitz}} \bibnamefont{and}
  \bibinfo{author}{\bibfnamefont{M.}~\bibnamefont{Esposito}},
  \bibinfo{journal}{Phys. Rev. X} \textbf{\bibinfo{volume}{4}},
  \bibinfo{pages}{031015} (\bibinfo{year}{2014}).

\bibitem[{\citenamefont{Amari}(2016)}]{Amari2016}
\bibinfo{author}{\bibfnamefont{S.-I.} \bibnamefont{Amari}},
  \emph{\bibinfo{title}{{Information geometry and its applications}}}
  (\bibinfo{publisher}{Springer, Tokyo}, \bibinfo{year}{2016}),
  \bibinfo{edition}{1st} ed.

\bibitem[{\citenamefont{Nielsen}(2020)}]{Nielsen2020}
\bibinfo{author}{\bibfnamefont{F.}~\bibnamefont{Nielsen}},
  \bibinfo{journal}{Entropy} \textbf{\bibinfo{volume}{22}},
  \bibinfo{pages}{1100} (\bibinfo{year}{2020}).

\bibitem[{\citenamefont{Hartich et~al.}(2014)\citenamefont{Hartich, Barato, and
  Seifert}}]{Hartich2014}
\bibinfo{author}{\bibfnamefont{D.}~\bibnamefont{Hartich}},
  \bibinfo{author}{\bibfnamefont{A.}~\bibnamefont{Barato}}, \bibnamefont{and}
  \bibinfo{author}{\bibfnamefont{U.}~\bibnamefont{Seifert}},
  \bibinfo{journal}{J. Stat. Mech Theory Exp.} \textbf{\bibinfo{volume}{2014}},
  \bibinfo{pages}{02016} (\bibinfo{year}{2014}).

\bibitem[{\citenamefont{Barato et~al.}(2013)\citenamefont{Barato, Hartich, and
  Seifert}}]{Barato2013}
\bibinfo{author}{\bibfnamefont{A.~C.} \bibnamefont{Barato}},
  \bibinfo{author}{\bibfnamefont{D.}~\bibnamefont{Hartich}}, \bibnamefont{and}
  \bibinfo{author}{\bibfnamefont{U.}~\bibnamefont{Seifert}},
  \bibinfo{journal}{J Stat Phys} \textbf{\bibinfo{volume}{153}}
  (\bibinfo{year}{2013}).

\bibitem[{\citenamefont{Chetrite et~al.}(2019)\citenamefont{Chetrite,
  Rosinberg, Sagawa, and Tarjus}}]{Chetrite2019}
\bibinfo{author}{\bibfnamefont{R.}~\bibnamefont{Chetrite}},
  \bibinfo{author}{\bibfnamefont{M.~L.} \bibnamefont{Rosinberg}},
  \bibinfo{author}{\bibfnamefont{T.}~\bibnamefont{Sagawa}}, \bibnamefont{and}
  \bibinfo{author}{\bibfnamefont{G.}~\bibnamefont{Tarjus}},
  \bibinfo{journal}{J. Stat. Mech.} \textbf{\bibinfo{volume}{21}},
  \bibinfo{pages}{114002} (\bibinfo{year}{2019}).

\bibitem[{\citenamefont{Li and Ma}(2016)}]{Li2016}
\bibinfo{author}{\bibfnamefont{W.}~\bibnamefont{Li}} \bibnamefont{and}
  \bibinfo{author}{\bibfnamefont{A.}~\bibnamefont{Ma}}, \bibinfo{journal}{J.
  Chem. Phys.} \textbf{\bibinfo{volume}{144}}, \bibinfo{pages}{114103}
  (\bibinfo{year}{2016}).

\bibitem[{\citenamefont{Johnson and Hummer}(2012)}]{Johnson2012}
\bibinfo{author}{\bibfnamefont{M.~E.} \bibnamefont{Johnson}} \bibnamefont{and}
  \bibinfo{author}{\bibfnamefont{G.}~\bibnamefont{Hummer}},
  \bibinfo{journal}{J. Phys. Chem. B} \textbf{\bibinfo{volume}{116}},
  \bibinfo{pages}{8573} (\bibinfo{year}{2012}).

\bibitem[{\citenamefont{Bolhuis et~al.}(2002)\citenamefont{Bolhuis, Chandler,
  Dellago, and Geissler}}]{Bolhuis2002}
\bibinfo{author}{\bibfnamefont{P.~G.} \bibnamefont{Bolhuis}},
  \bibinfo{author}{\bibfnamefont{D.}~\bibnamefont{Chandler}},
  \bibinfo{author}{\bibfnamefont{C.}~\bibnamefont{Dellago}}, \bibnamefont{and}
  \bibinfo{author}{\bibfnamefont{P.~L.} \bibnamefont{Geissler}},
  \bibinfo{journal}{Annu. Rev. Phys. Chem} \textbf{\bibinfo{volume}{53}},
  \bibinfo{pages}{291} (\bibinfo{year}{2002}).

\bibitem[{\citenamefont{Van~Erp et~al.}(2003)\citenamefont{Van~Erp, Moroni, and
  Bolhuis}}]{vanErp2003}
\bibinfo{author}{\bibfnamefont{T.~S.} \bibnamefont{Van~Erp}},
  \bibinfo{author}{\bibfnamefont{D.}~\bibnamefont{Moroni}}, \bibnamefont{and}
  \bibinfo{author}{\bibfnamefont{P.~G.} \bibnamefont{Bolhuis}},
  \bibinfo{journal}{J. Chem. Phys.} \textbf{\bibinfo{volume}{118}},
  \bibinfo{pages}{6617} (\bibinfo{year}{2003}).

\bibitem[{\citenamefont{Allen et~al.}(2005)\citenamefont{Allen, Warren, and
  Ten~Wolde}}]{Allen2005}
\bibinfo{author}{\bibfnamefont{R.~J.} \bibnamefont{Allen}},
  \bibinfo{author}{\bibfnamefont{P.~B.} \bibnamefont{Warren}},
  \bibnamefont{and} \bibinfo{author}{\bibfnamefont{P.~R.}
  \bibnamefont{Ten~Wolde}}, \bibinfo{journal}{Phys. Rev. Lett.}
  \textbf{\bibinfo{volume}{94}}, \bibinfo{pages}{018104}
  (\bibinfo{year}{2005}).

\bibitem[{\citenamefont{Faradjian and Elber}(2004)}]{Elber2004}
\bibinfo{author}{\bibfnamefont{A.~K.} \bibnamefont{Faradjian}}
  \bibnamefont{and} \bibinfo{author}{\bibfnamefont{R.}~\bibnamefont{Elber}},
  \bibinfo{journal}{J. Chem. Phys.} \textbf{\bibinfo{volume}{120}},
  \bibinfo{pages}{10880} (\bibinfo{year}{2004}).

\bibitem[{\citenamefont{Seifert}(2019)}]{Seifert2019}
\bibinfo{author}{\bibfnamefont{U.}~\bibnamefont{Seifert}},
  \bibinfo{journal}{Annu. Rev. Condens. Matter Phys.}
  \textbf{\bibinfo{volume}{10}}, \bibinfo{pages}{171} (\bibinfo{year}{2019}).

\bibitem[{\citenamefont{Li et~al.}(2019{\natexlab{a}})\citenamefont{Li,
  Gingrich, and Fakhri}}]{Li2019a}
\bibinfo{author}{\bibfnamefont{J.~M.} \bibnamefont{Li},
  \bibfnamefont{Junang~andHorowitz}}, \bibinfo{author}{\bibfnamefont{T.~R.}
  \bibnamefont{Gingrich}}, \bibnamefont{and}
  \bibinfo{author}{\bibfnamefont{N.}~\bibnamefont{Fakhri}},
  \bibinfo{journal}{Nature Communications} \textbf{\bibinfo{volume}{10}}
  (\bibinfo{year}{2019}{\natexlab{a}}).

\bibitem[{\citenamefont{Skinner and Dunkel}(2021)}]{Skinner2021}
\bibinfo{author}{\bibfnamefont{D.~J.} \bibnamefont{Skinner}} \bibnamefont{and}
  \bibinfo{author}{\bibfnamefont{J.}~\bibnamefont{Dunkel}},
  \bibinfo{journal}{PNAS} \textbf{\bibinfo{volume}{118}}
  (\bibinfo{year}{2021}).

\bibitem[{\citenamefont{Gnesotto et~al.}(2020)\citenamefont{Gnesotto, Gradziuk,
  Ronceray, and Broedersz}}]{Gnesotto2020}
\bibinfo{author}{\bibfnamefont{F.~S.} \bibnamefont{Gnesotto}},
  \bibinfo{author}{\bibfnamefont{G.}~\bibnamefont{Gradziuk}},
  \bibinfo{author}{\bibfnamefont{P.}~\bibnamefont{Ronceray}}, \bibnamefont{and}
  \bibinfo{author}{\bibfnamefont{C.~P.} \bibnamefont{Broedersz}},
  \bibinfo{journal}{Nat. Commun.} \textbf{\bibinfo{volume}{11}},
  \bibinfo{pages}{5378} (\bibinfo{year}{2020}).

\bibitem[{\citenamefont{Khoo et~al.}(2019)\citenamefont{Khoo, Lu, and
  Ying}}]{Khoo2019}
\bibinfo{author}{\bibfnamefont{Y.}~\bibnamefont{Khoo}},
  \bibinfo{author}{\bibfnamefont{J.}~\bibnamefont{Lu}}, \bibnamefont{and}
  \bibinfo{author}{\bibfnamefont{L.}~\bibnamefont{Ying}},
  \bibinfo{journal}{Res. Math. Sci.} \textbf{\bibinfo{volume}{6}},
  \bibinfo{pages}{1} (\bibinfo{year}{2019}).

\bibitem[{\citenamefont{Li et~al.}(2019{\natexlab{b}})\citenamefont{Li, Lin,
  and Ren}}]{Li2019}
\bibinfo{author}{\bibfnamefont{Q.}~\bibnamefont{Li}},
  \bibinfo{author}{\bibfnamefont{B.}~\bibnamefont{Lin}}, \bibnamefont{and}
  \bibinfo{author}{\bibfnamefont{W.}~\bibnamefont{Ren}}, \bibinfo{journal}{J.
  Chem. Phys.} \textbf{\bibinfo{volume}{151}}, \bibinfo{pages}{54112}
  (\bibinfo{year}{2019}{\natexlab{b}}).

\bibitem[{\citenamefont{Rotskoff et~al.}()\citenamefont{Rotskoff, Mitchell, and
  Vanden-Eijnden}}]{Rotskoff2021}
\bibinfo{author}{\bibfnamefont{G.~M.} \bibnamefont{Rotskoff}},
  \bibinfo{author}{\bibfnamefont{A.~R.} \bibnamefont{Mitchell}},
  \bibnamefont{and}
  \bibinfo{author}{\bibfnamefont{E.}~\bibnamefont{Vanden-Eijnden}},
  \bibinfo{note}{arXiv:2008.06334v2}.

\bibitem[{\citenamefont{Ma and Dinner}(2005)}]{Ma2005}
\bibinfo{author}{\bibfnamefont{A.}~\bibnamefont{Ma}} \bibnamefont{and}
  \bibinfo{author}{\bibfnamefont{A.~R.} \bibnamefont{Dinner}},
  \bibinfo{journal}{J. Phys. Chem. B} \textbf{\bibinfo{volume}{109}},
  \bibinfo{pages}{6769} (\bibinfo{year}{2005}).

\bibitem[{\citenamefont{Wang and Tiwary}(2021)}]{Wang2021}
\bibinfo{author}{\bibfnamefont{Y.}~\bibnamefont{Wang}} \bibnamefont{and}
  \bibinfo{author}{\bibfnamefont{P.}~\bibnamefont{Tiwary}},
  \bibinfo{journal}{J. Chem. Phys.} \textbf{\bibinfo{volume}{154}},
  \bibinfo{pages}{134111} (\bibinfo{year}{2021}).

\bibitem[{\citenamefont{Wang et~al.}(2020)\citenamefont{Wang, Ribeiro, and
  Tiwary}}]{Wang2020}
\bibinfo{author}{\bibfnamefont{Y.}~\bibnamefont{Wang}},
  \bibinfo{author}{\bibfnamefont{J.~M.~L.} \bibnamefont{Ribeiro}},
  \bibnamefont{and} \bibinfo{author}{\bibfnamefont{P.}~\bibnamefont{Tiwary}},
  \bibinfo{journal}{Curr. Opin. Struct. Biol.} \textbf{\bibinfo{volume}{61}},
  \bibinfo{pages}{139} (\bibinfo{year}{2020}).

\bibitem[{\citenamefont{Ito}(2018)}]{Ito2018}
\bibinfo{author}{\bibfnamefont{S.}~\bibnamefont{Ito}}, \bibinfo{journal}{Phys.
  Rev. Lett.} \textbf{\bibinfo{volume}{121}}, \bibinfo{pages}{30605}
  (\bibinfo{year}{2018}).

\bibitem[{\citenamefont{Ruppeiner}(1979)}]{Ruppeiner1979}
\bibinfo{author}{\bibfnamefont{G.}~\bibnamefont{Ruppeiner}},
  \bibinfo{journal}{Phys. Rev. A} \textbf{\bibinfo{volume}{20}},
  \bibinfo{pages}{1608} (\bibinfo{year}{1979}).

\bibitem[{\citenamefont{Crooks}(2007)}]{Crooks2007}
\bibinfo{author}{\bibfnamefont{G.~E.} \bibnamefont{Crooks}},
  \bibinfo{journal}{Phys. Rev. Lett.} \textbf{\bibinfo{volume}{99}},
  \bibinfo{pages}{100602} (\bibinfo{year}{2007}).

\bibitem[{\citenamefont{Metropolis et~al.}(1953)\citenamefont{Metropolis,
  Rosenbluth, Rosenbluth, Teller, and Teller}}]{Metropolis1953}
\bibinfo{author}{\bibfnamefont{N.}~\bibnamefont{Metropolis}},
  \bibinfo{author}{\bibfnamefont{A.~W.} \bibnamefont{Rosenbluth}},
  \bibinfo{author}{\bibfnamefont{M.~N.} \bibnamefont{Rosenbluth}},
  \bibinfo{author}{\bibfnamefont{A.~H.} \bibnamefont{Teller}},
  \bibnamefont{and} \bibinfo{author}{\bibfnamefont{E.}~\bibnamefont{Teller}},
  \bibinfo{journal}{J. Chem. Phys.} \textbf{\bibinfo{volume}{21}},
  \bibinfo{pages}{1087} (\bibinfo{year}{1953}).

\end{thebibliography}

\onecolumngrid
\clearpage
\begin{center}
	\textbf{\large Supplemental Material for ``Information Thermodynamics of the Transition-Path Ensemble''}
\end{center}
\setcounter{equation}{0}
\setcounter{figure}{0}
\setcounter{table}{0}
\setcounter{page}{1}
\makeatletter
\renewcommand{\theequation}{S\arabic{equation}}
\renewcommand{\thefigure}{S\arabic{figure}}

\section{Joint transition rates}
\label{appendix_transition_rates}

Following \cite{Vanden-Eijnden2014}, the transition probability over time $\md t$ for a $\st' \to \st$ transition given the trajectory remains in subensemble $\subens$ is
\begin{subequations}
\begin{align} 
    T^{\subens=\subens'}_{\st \st'} \md t &= p(\st,\out,\org|\st',\out',\org') \label{eq:write_trans_prob} \\
    &= p(\out,\org|\st,\st',\out',\org') \ p(\st|\st',\out',\org') \label{eq:fixedSuben_splitJoint} \\
    &= p(\st|\st',\out',\org')  \label{eq:out_org_fixed} \\
    &= p(\st|\st',\out') \label{eq:elim_origin} \\
    &= \frac{p(\st,\out'|\st')}{p(\out'|\st')} \label{eq:write_joint_marg} \\
    &= \frac{p(\out'|\st)\, p(\st|\st')}{p(\out'|\st')}  \label{eq:separate_committor} \>.
\end{align}
\end{subequations}
\eqref{eq:fixedSuben_splitJoint} splits the joint probability into conditional and marginal probabilities.
In \eqref{eq:out_org_fixed}, we recognize that $p(\out,\org|\st,\st',\out',\org')=1$ when the subensemble doesn't change. In \eqref{eq:elim_origin}, we use the Markov property to eliminate the dependence of the next state on trajectory origin. In \eqref{eq:write_joint_marg}, we express the conditional probability as the ratio of joint and marginal probabilities. In \eqref{eq:separate_committor}, we recognize that the trajectory outcome depends only on the most recent state $\st$ and drop the dependency on $\st'$. Finally, we recall that for these transitions $\out=\out'$ and re-express the transition probability as a transition rate,
\begin{align}
T^{\subens=\subens'}_{\st \st'} = \frac{p(\out|\st)}{p(\out|\st')} T_{\st \st'} \label{eq:write_eqm_rate} \>.
\end{align}

We similarly derive transition rates for the four sets of transitions that change subensemble, where either the trajectory origin or outcome changes while the other is constant. The transition probability in time $\md t$ for a $\st' \to \st$ transition out of $A$ where the subensemble changes from $\subens'=(A,A) \to \subens=(A,B)$ is:
\begin{subequations}
\begin{align}
    T^{\subens=(A,B), \subens'=(A,A)}_{\st \st'} \md t &= p(\st,\Out=B,\Org=A|\st',\Out'=A,\Org'=A)  \label{eq:change_out_prob} \\
    &= p(\Out=B,\Org=A|\st,\st',\Out'=A,\Org'=A) \ p(\st|\st',\Out'=A,\Org'=A)  \label{eq:AAtoAB_jointToMargCond} \\
    &= p(\Out=B|\st,\st',\Out'=A,\Org'=A) \ p(\st|\st',\Out'=A,\Org'=A)  \label{eq:conditional_independence} \\
    &= p(\Out=B|\st) \ p(\st|\st') \label{eq:simplify_out_change} \ .
\end{align}
\end{subequations}
\eqref{eq:conditional_independence} uses the fact that $\Org=A$ when $\st' \in A$. \eqref{eq:simplify_out_change} uses the Markov property to simplify the conditional distributions. We then re-express the transition probability as a transition rate
\begin{align}
    T^{\subens=(A,B), \subens'=(A,A)}_{\st \st'} = p(\Out=B|\st) \, T_{\st \st'} \>.
\end{align}
Similarly, the transition rate for a $\st' \to \st$ transition where the subensemble outcome changes from $\subens'=(B,B) \to \subens'=(B,A)$ is 
\begin{align}
    T^{\subens=(B,A), \subens'=(B,B)}_{\st \st'} = p(\Out=A|\st) \, T_{\st \st'} \>.
\end{align}

The trajectory origin changes when the system finishes a transition path at the boundary of $A$ or $B$. The probability for a $\st' \to \st$ transition into $A$ where the subensemble changes from $\subens'=(B,A) \to \subens=(A,A)$ is:
\begin{subequations}
\begin{align}
    T^{\subens=(A,A), \subens'=(B,A)}_{\st \st'} \md t &= p(\st,\Out=A,\Org=A|\st',\Out'=A,\Org'=B) \\
    &= p(\Out=A,\Org=A|\st,\st',\Out'=A,\Org'=B) \ p(\st|\st',\Out'=A,\Org'=B) \label{eq:origin_cond_ind}  \\
    &= p(\st|\st',\Out'=A) \label{eq:origin_elimA}   \\
    &= \frac{p(\st,\Out'=A|\st')}{p(\Out'=A|\st')} \\
    &= \frac{p(\Out'=A|\st,\st')\, p(\st|\st')}{p(\Out'=A|\st')} \\
    &= \frac{p(\st|\st')}{p(\Out'=A|\st')} \label{eq:origin_simplify} \>.
\end{align}
\end{subequations}
\eqref{eq:origin_elimA} recognizes that $\Out=A$ and $\Org=A$ for $\st \in A$ and uses the Markov property to eliminate dependence on $\Org'$ in $p(\st|\st',\Out'=A,\Org'=A)$. \eqref{eq:origin_simplify} uses $\Out'=A$ for $\st \in A$. Finally, we re-express the transition probability as the transition rate
\begin{align}
    T^{\subens=(A,A), \subens'=(B,A)}_{\st \st'} = \frac{1}{p(\Out=A|\st')} T_{\st \st'} \>,
\end{align}
and similarly derive the transition rate for a $\st' \to \st$ transition where the system enters $B$ and finishes a forward TPE ($\subens'=(A,B) \to \subens=(B,B)$) as
\begin{align}
    T^{\subens=(B,B), \subens'=(A,B)}_{\st \st'} = \frac{1}{p(\Out=B|\st')} T_{\st \st'} \>.
\end{align}

These rates are explicitly written for each transition in \eqref{eq:nuS}, and when averaged over all such transitions yield the unidirectional probability flux $\nu_{\Subens}$ between trajectory subensembles~\eqref{eq:rxn_parameters},
\begin{subequations} \label{eq:uni_prob_flux}
\begin{align}
    \nu_{\Subens} &= \sum_{\st \notin A, \st' \in A} p(\Out=B|\st) \, T_{\st \st'} \pi(\st') \label{eq:uni_prob_flux_AB} \\
    &= \sum_{\st \notin B, \st' \in B} p(\Out=A|\st) \, T_{\st \st'} \pi(\st') \\
    &= \sum_{\st \in A, \st' \notin A} T_{\st \st'} \pi(\st') \, p(\Org=B|\st')  \\
    &= \sum_{\st \in B, \st' \notin B} T_{\st \st'} \pi(\st') \, p(\Org=A|\st')  \>.
\end{align}
\end{subequations}
Note that each RHS of \eqref{eq:uni_prob_flux} has an implicit conditional probability for the other element of the trajectory subsensemble, each of which equals unity on the relevant system subspace, e.g., $p(\Org=A|\st')=1$ for $\st'\in A$ in \eqref{eq:uni_prob_flux_AB}.

\section{Entropy production for joint dynamics in $(\St,\Subens)$}
\label{appendix_entropy_transitions}

We decompose the change in joint entropy~\eqref{eq:joint_entropy_production} into three terms~\cite{Busiello2020}
\begin{subequations} \label{eq:decompose_entropy}
\begin{align}
    0 &= \md_t H(\St,\Subens) = 
    \underbrace{\sum_{\st,\st', \subens} T^{\subens}_{\st \st'} p(\st',\subens) \ln \frac{T^{\subens}_{\st \st'} p(\st',\subens)}{T^{\subens}_{\st' \st} p(\st,\subens)} }_{\dot{H}^{\rm{irr}}(\St,\Subens)} 
    - \underbrace{\sum_{\st,\st', \subens} T^{\subens}_{\st \st'} p(\st',\subens) \ln \frac{T^{\subens}_{\st \st'}}{T^{\subens}_{\st' \st}}}_{\dot{H}^{\rm{env}}(\St,\Subens)} 
    + \underbrace{\sum_{\st,\st', \subens\neq\subens'} T^{\subens \subens'}_{\st \st'} p(\st',\subens') \ln \frac{p(\st',\subens')}{p(\st,\subens)}}_{\dot{H}^{\rm{sub}}(\St,\Subens)}\>, 
\end{align}
\end{subequations}
where $\dot{H}^{\rm{irr}}(\St,\Subens)$ is the irreversible entropy production and $\dot{H}^{\rm{env}}(\St,\Subens)$ the environmental entropy change for transitions that do not change the subensemble, and $\dot{H}^{\rm{sub}}(\St,\Subens)$ is the change in joint entropy due to transitions that change the subensemble. 

The transitions that change the trajectory subensemble do not change joint entropy:
\begin{subequations} \label{eq:uni_entropy}
\begin{align}
    \dot{H}&^{\rm{sub}}(\St,\Subens) \nonumber \\
    &= \sum_{\st \notin A, \st' \in A} T_{\st' \st} \pi(\st) p(\Org = B|\st) \ln \frac{p(\st,\Subens=(B,A))}{\pi(\st')} 
    -\sum_{\st \notin A, \st' \in A} T_{\st \st'} \pi(\st') p(\Out = B|\st) \ln \frac{p(\st,\Subens=(A,B))}{\pi(\st')} \label{eq:HdotUniExpanded} \\
    & \quad + \sum_{\st \notin B, \st' \in B} T_{\st' \st} \pi(\st) p(\Org = A|\st) \ln \frac{p(\st,\Subens=(A,B))}{\pi(\st')} 
    - \sum_{\st \notin B, \st' \in B} T_{\st \st'} \pi(\st') p(\Out = A|\st) \ln \frac{p(\st,\Subens=(B,A))}{\pi(\st')} \nonumber \\
    &= 0 \>.
\end{align}
\end{subequations}
In \eqref{eq:HdotUniExpanded}, the first and second terms and the third and fourth terms cancel because of the equilibrium relationships $p(\Out=A|\st)=p(\Org=A|\st)$, $p(\Out=B|\st)=p(\Org=B|\st)$, and $p(\phi,\Subens=(B,A))=p(\phi,\Subens=(A,B))$.

We express the irreversible entropy production $\dot{H}^{\rm{irr}}(\St,\Subens)$ in terms of the irreversible entropy production of dynamics given fixed subensemble $\subens$, $\dot{\Sigma}_{\subens}$~\cite{Esposito2012}:
\begin{subequations} \label{eq:irr_entropy}
\begin{align}
    \dot{H}^{\rm{irr}}(\St,\Subens) &= \sum_{\subens} p(\subens) \sum_{\st,\st'} T^{\subens}_{\st \st'} p(\st'|\subens) \ln \frac{T^{\subens}_{\st \st'} p(\st'|\subens)}{T^{\subens}_{\st' \st} p(\st|\subens)} \\
    &= \sum_{\subens} p(\subens) \dot{\Sigma}_{\subens} \\
    &\equiv \langle \dot{\Sigma} \rangle \ .
\end{align}
\end{subequations}

Finally, the environmental entropy change is
\begin{subequations} \label{eq:env_entropy}
\begin{align}
    \dot{H}^{\rm{env}}(\St,\Subens) &= 2 \sum_{\st,\st', \subens} T^{\subens}_{\st \st'} p(\st',\subens) \ln \frac{p(\out|\st)}{p(\out|\st')} \\
    &= 2 \sum_{\st,\st', \out} T^{\out}_{\st \st'} p(\st',\out) \ln \frac{p(\out|\st)}{p(\out|\st')} \label{eq:sum_origin} \\
    &= 2 \dot{I}^{\St}(\Out;\St) \>,
\end{align}
\end{subequations}
where we get \eqref{eq:sum_origin} by summing over the trajectory origin.

\section{Mutual information rates between system state and either trajectory outcome or trajectory origin have equal magnitude} \label{appendix_MI_entropy}

The rate of change in mutual information between system state and trajectory \emph{outcome} due to $\St$ dynamics is of equal magnitude and opposite sign from the rate of change in mutual information between system state and trajectory \emph{origin} due to $\St$ dynamics:
\begin{subequations}
\begin{align}
    \dot{I}^{\St}(\Out;\St) &= \sum_{\st,\st',\out} T^{\out}_{\st \st'} p(\st',\out) \ln \frac{p(\out|\st)}{p(\out|\st')} \\
    &= \sum_{\st,\st',\out} T_{\st \st'} \pi(\st') p(\out|\st) \ln \frac{p(\out|\st)}{p(\out|\st')} \label{eq:write_out_flux} \\
    &= \sum_{\st,\st',\org} T_{\st \st'} \pi(\st') p(\org|\st) \ln \frac{p(\org|\st)}{p(\org|\st')} \label{eq:replace_out_org} \\
    &= \sum_{\st,\st',\org} T_{\st' \st} \pi(\st) p(\org|\st) \ln \frac{p(\org|\st)}{p(\org|\st')} \label{eq:relate_eqm_flux} \\
    &= -\sum_{\st,\st',\org} T_{\st' \st} p(\st,\org) \ln \frac{p(\org|\st')}{p(\org|\st)}  \\
    &= -\dot{I}^{\St}(\Org;\St) \>.
\end{align}
\end{subequations}
In \eqref{eq:write_out_flux} we expand the joint transition rate using \eqref{eq:TPE_transition_rates}; \eqref{eq:replace_out_org} relates outcome and origin probabilities using equilibrium relations $p(\Out=A|\st)=p(\Org=A|\st)$, $p(\Out=B|\st)=p(\Org=B|\st)$; and \eqref{eq:relate_eqm_flux} uses detailed balance, $T_{\st \st'} \pi(\st')=T_{\st' \st} \pi(\st)$.

\section{Thermodynamic metric} \label{appendix_TD_metric}

Here we assume that the state space $\St$ is continuous, and the master equation $\md_t p(\st) = \sum_{\st'} T_{\st \st'} p(\st')$ represents a discrete approximation of its dynamics. We rearrange the difference in mutual information rates~\eqref{eq:entropy_info_difference} to obtain the transition-weighted relative entropy $D[p(\subens|\st')||p(\subens|\st)]\equiv \sum_{\subens} p(\subens|\st') \ln p(\subens|\st')/p(\subens|\st)$ between the conditional subensemble distributions $p(\subens|\st')$ and $p(\subens|\st)$ before and after the transition, respectively, then expand in small state changes $\st-\st'$: 
\begin{subequations}
\begin{align} 
p_{\TPE} \dot{\Sigma}_{\TPE} &= \dot{I}^{\St}(\Out;\St)-\dot{I}^{\St}(\Org;\St) \\
    &= \sum_{\st,\st',\out} 
    T_{\st \st'} \pi(\st') p(\out|\st) \ln \frac{p(\out|\st)}{p(\out|\st')} 
    - \sum_{\st,\st',\org} T_{\st' \st} \pi(\st) p(\org|\st) \ln \frac{p(\org|\st')}{p(\org|\st)} 
    \label{eq:write_info_rates} \\
    &= \sum_{\st,\st',\out,\org} 
    T_{\st \st'} \pi(\st') p(\out|\st)p(\org|\st) \ln \frac{p(\out|\st)}{p(\out|\st')} 
    - \sum_{\st,\st',\out,\org} T_{\st' \st} \pi(\st) p(\org|\st) p(\out|\st) \ln \frac{p(\org|\st')}{p(\org|\st)} 
    \label{eq:multiply_unity} \\
    &= \sum_{\st,\st',\out,\org} 
    T_{\st \st'} \pi(\st') p(\out|\st) p(\org|\st) \ln \frac{p(\out|\st)p(\org|\st)}{p(\out|\st')p(\org|\st')} \label{eq:combine_terms} \\
    &= \sum_{\st,\st'} T_{\st \st'} \pi(\st') D[p(\subens|\st') || p(\subens|\st)] \\ 
    &\approx \sum_{\st,\st'} T_{\st \st'} \pi(\st') \tfrac{1}{2} \sum_{i,j} (\phi_i - \phi'_i) \mathcal{I}_{ij}(\st') (\phi_j-\phi'_j) \label{eq:rate_RC_distance} \> .
\end{align}
\end{subequations}
In \eqref{eq:multiply_unity}, we multiply each term by unity ($\sum_{\org} p(\org|\st)$ and $\sum_{\out} p(\out|\st)$ respectively), then use detailed balance ($T_{\st \st'} \pi(\st') = T_{\st' \st} \pi(\st)$) to combine terms in \eqref{eq:combine_terms}. $\phi_i$ is the $i$th component of the state-space vector and $\mathcal{I}_{ij}(\st)$ is the Fisher information of the trajectory outcome/origin distribution at state $\st$,
\begin{subequations} \label{eq:TPE_Fisher_info}
\begin{align}
\mathcal{I}_{ij}(\st) &\equiv \sum_{\subens} p \left(\subens|\st \right) \frac{\partial \ln p(\subens|\st)}{\partial \st_i} \frac{\partial \ln p(\subens|\st)}{\partial \st_j} \\
&= \sum_{\subens} \frac{1}{p\left(\subens|\st\right)} \frac{\partial p(\subens|\st)}{\partial \st_i} \frac{\partial p(\subens|\st)}{\partial \st_j} \label{eq:FI_sum} \\
&= \frac{1}{(1-q^+_{\st})q^+_{\st}} \frac{\partial (1-q^+_{\st})q^+_{\st}}{\partial \st_i} \frac{\partial (1-q^+_{\st})q^+_{\st}}{\partial \st_j} + \frac{1}{(1-q^+_{\st})^2} \frac{\partial (1-q^+_{\st})^2}{\partial \st_i} \frac{\partial (1-q^+_{\st})^2}{\partial \st_j} \nonumber \label{eq:write_terms} \\
&\quad + \frac{1}{(q^+_{\st})^2} \frac{\partial (q^+_{\st})^2}{\partial \st_i} \frac{\partial (q^+_{\st})^2}{\partial \st_j} + \frac{1}{q^+_{\st}(1-q^+_{\st})} \frac{\partial q^+_{\st}(1-q^+_{\st})}{\partial \st_i} \frac{\partial q^+_{\st}(1-q^+_{\st})}{\partial \st_j} \\
&=\left[ \frac{(1-2q^+_{\st})^2}{(1-q^+_{\st})q^+_{\st}} + \frac{4(1-q^+_{\st})^2}{(1-q^+_{\st})^2} + \frac{4(q^+_{\st})^2}{(q^+_{\st})^2} + \frac{(1-2 q^+_{\st})^2}{q^+_{\st}(1-q^+_{\st})} \right] \frac{\partial q^+_{\st}}{\partial \st_i} \frac{\partial q^+_{\st}}{\partial \st_j} \label{eq:collect_terms} \\
&= \frac{2}{q^+_{\st} \left(1-q^+_{\st}\right)} \frac{\partial q^+_{\st}}{\partial \st_i} \frac{\partial q^+_{\st}}{\partial \st_j} \label{eq:simplify_FI} \> .
\end{align}
\end{subequations}
In \eqref{eq:write_terms}, we write out each term from \eqref{eq:FI_sum} in terms of the forward committor $q^+_{\st}$. We use the chain rule to pull out a common factor in \eqref{eq:collect_terms} and simplify in \eqref{eq:simplify_FI}.

In information geometry~\cite{Amari2016,Nielsen2020}, Fisher information arises as a distance metric in the parameter space of a probability distribution. Here, we consider changes in conditional probability $p(\subens|\st)$ as the system evolves, where the system state parameterizes the conditional probability distribution through the committor $q^+_{\st}$. When the system is in $A$ ($B$), there is no uncertainty about trajectory outcome $\Out$ and origin $\Org$, so the conditional probability $p(\subens|\st \in A)$ ($p(\subens|\st \in B)$) is unity for $\subens=(A,A)$ ($\subens=(B,B)$) and zero for all other subensembles. As the system evolves, the distribution $p(\subens|\st)$ corresponding to the current system state changes, and the information-geometric distance between successive distributions is quantified by the Fisher information metric, with square line element~\cite{Ito2018} 
\begin{align}
\md \ell_{\st \st'}^2 \equiv \tfrac{1}{2} \sum_{i,j} (\phi_i - \phi'_i) \mathcal{I}_{ij}(\st') (\phi_j-\phi'_j) \>.
\end{align}

Equation~\eqref{eq:rate_RC_distance} is then the rate of mean square distance accumulated by the system evolving at equilibrium, 
\begin{align}
\frac{\langle \md \ell^2
\rangle}{\md t} = \sum_{\st,\st'} T_{\st \st'} \pi(\st') \md \ell_{\st \st'}^2 \>. 
\end{align} 
Multiplying by the mean round-trip time $\tau_A+\tau_B = (\nu_{\subens})^{-1}$ through the subensembles (the sum of mean first-passage times from $A$ to $B$ and $B$ to $A$~\cite{Berezhkovskii2019}), we obtain the squared reaction-coordinate length $\mathcal{L}_{AB}^2$ as the mean square metric distance for a round trip $A \to B \to A$:
\begin{subequations} 
\begin{align}
\mathcal{L}_{AB}^2 &\equiv (\tau_A+\tau_B) \frac{\langle \md \ell^2 \rangle}{\md t} \\ 
&\approx (\tau_A+\tau_B) 2 p_{\TPE} \dot{\Sigma}_{\TPE} \\ 
&= 2 \tau_{\TPE} \dot{\Sigma}_{\TPE} \\
\dot{\Sigma}_{\TPE} &\approx \frac{\mathcal{L}_{AB}^2}{2 \tau_{\TPE}} \>,
\end{align}
\end{subequations}
for mean transition-path duration $\tau_{\TPE} = (\tau_A+\tau_B)p_{\TPE}$. The reaction-coordinate length $\mathcal{L}_{AB}$ roughly quantifies the mean number of fluctuations~\cite{Ruppeiner1979,Crooks2007} required for the system to complete a round trip.

\section{Computational details}
\label{appendix_bistable}

The bistable energy potential is separable into terms only depending on the reaction coordinate $r$ and on the bath mode $b$:
\begin{align} \label{eq:bistable_energy}
E(r,b) = -\kT \ln \left[e^{-\tfrac{1}{2} \beta k_{\rm{m}} (r+r_{\rm{m}})^2}+e^{-\tfrac{1}{2} \beta k_{\rm{m}} (r-r_{\rm{m}})^2}\right] + \tfrac{1}{2}k b^2 \>,
\end{align}
where $r_{\rm{m}}=1$ defines the locations of the energy minima. $k_{\rm{m}}$ is the landscape curvature near those energy minima, chosen such that the energy barrier $E(0,0)-E(r_{\rm{m}},0)$ is $4 \kT$. 

We represent the system state with orthogonal coordinates $(x,y)$, related to $(r,b)$ by rotational angle $\theta$:
\begin{subequations}
\begin{align}
    r &= x \cos \theta + y \sin \theta \\
    b &= -x \cos \theta + y \sin \theta \>.
\end{align}
\end{subequations}
We discretize the state space $(x,y)$ so that $\md x = \md y=0.04 r_{\rm{m}}$, and dynamically evolve bipartite dynamics using the master equation. The transition rate is $T_{x x', y y'}=\Gamma * {\rm{ min}} \left[ 1,e^{-\beta \Delta E} \right]$ with diffusion prefactor $\Gamma=0.1 \, \md t^{-1}$ and energy change $\Delta E = E(x,y)-E(x',y')$~\cite{Metropolis1953}. We solve the committor on the discrete state space using the recursion relations~\cite{Metzner2009}
\begin{subequations} \label{eq:fwd_committor_recursion}
\begin{align}
    q^+_{\st} &= \md t \sum_{\st'} T_{\st' \st} q^{\rm{+}}_{\st'} \\
    q^-_{\st} &= \md t \sum_{\st'} T_{\st \st'} \frac{\pi(\st')}{\pi(\st)} q^-_{\st'} \>.
\end{align}
\end{subequations}
Mesostates are defined by $A = \{ x,y \, | \, r(x,y) \le -r_m \}$ and $B = \{ x,y \, | \, r(x,y) \ge r_m \}$, so that the committor is independent of $b$. We calculate the TPE entropy production from \eqref{eq:bipartite_dissipation}.

\end{document}